\documentclass[11pt]{article}
\usepackage{rotating}
\usepackage{pbox}
\usepackage{amsmath}
\usepackage{amssymb}
\usepackage{amsfonts} 
\newcommand{\be}{\begin{equation}}
\newcommand{\ba}{\begin{eqnarray}}
\newcommand{\ee}{\end{equation}}
\newcommand{\ea}{\end{eqnarray}}
\newcommand{\cosec} { {\rm cosec}}

\begin{document}

\title{A class of exactly solvable rationally extended Calogero-Wolfes type $3$-body problems}

\author{Nisha Kumari$^{a}$\footnote{e-mail address: nishaism0086@gmail.com (N.K)}, Rajesh Kumar Yadav$^{b}$\footnote{e-mail address: rajeshastrophysics@gmail.com (R.K.Y)}, Avinash Khare$^{c}$\footnote {e-mail address: khare@physics.univpune.ac.in (A.K)} and \\
 Bhabani Prasad Mandal$^{a}$\footnote{e-mail address: bhabani.mandal@gmail.com (B.P.M).}}
 \maketitle
{$~^a$ Department of Physics, Banaras Hindu University, Varanasi-221005, INDIA.\\
$~^b$ Department of Physics, S. P. College, S. K. M. University, Dumka-814101, INDIA.\\ 
$~^c$Department of Physics, Savitribai Phule Pune University, Pune-411007, INDIA.}

\begin{abstract}

In this work, we start from the well known Calogero-Wolfes type $3$-body 
problems on a line and construct the corresponding exactly solvable rationally 
extended $3$-body potentials. In particular, we obtain 
the corresponding energy eigenvalues and eigenfunctions which are in terms of 
the product of $X_m$ Laguerre and $X_p$ Jacobi exceptional orthogonal 
polynomials where both $m,p = 1,2,3,...$. 
\end{abstract}

\section{Introduction}

There are very few exactly solvable quantum many body problems even in one
dimension. 
One of the first successful attempt was made by Calogero \cite{cal_69} in 
$1969$ who gave exact solution for the problem of three particles  interacting 
pairwise by inverse cube forces (i.e. inverse square potential) 
in addition to linear forces (i.e. harmonic potential). 
Thereafter Calogero and Sutherland  obtained 
the solutions of more general $N$-body problems in one dimension 
\cite{cal_71,suth_71}. Other one dimensional 
many body problems of the Calogero and the Sutherland types and their 
applications are also discussed  in Refs. 
\cite{cal_mar_74, ols_pere, poly_92, shas_88, hald_88, frahm_93}. 

In $1974$ Wolfes \cite{wolf_74} showed that the
Calogero approach could be generalized and one can obtain exact solution 
of new three-body problems in one dimension. Later on  Khare and Bhaduri
\cite{kh_bh_93} extended the Calogero-Wolfes approach and obtained exact 
solutions of a number of three-body potentials in one dimension by using the 
ideas of supersymmetric quantum mechanics \cite{cks}. After the development 
of $PT$ (combined parity and time reversal) symmetric
quantum mechanics \cite{bender}, people also considered the complex but
$PT$-invariant extension of the Calogero model and showed that in this case
too one has real bound state energy eigenvalue spectrum \cite{bpm, bpm1,bpm2,mz_08}.
It may be noted that the corresponding bound state eigenfunctions 
are associated with the well known classical orthogonal polynomials.

After the recent discovery of two new orthogonal polynomials namely the 
exceptional $X_1$ Laguerre and $X_1$ Jacobi (or more general 
$X_m$ Laguerre and $X_p$ Jacobi respectively)
polynomials \cite{dnr1,dnr2,xm1,xm2}, a number of one body exactly solvable 
potentials  \cite{que,bqr,os,hos,hs,gom,qu,yg1,dim} as well as Calogero type many particle potentials \cite{bpm5} have been extended rationally 
whose solutions are obtained in terms of these exceptional orthogonal 
polynomials (EOPs).   Many interesting 
properties of these extended potentials have been studied  \cite{pdm,nfold2,
fplank,codext1,codext2,qscat,scatpt, tdse,gtextd,para_sym}. It is then worth enquiring
if one can construct new Calogero-Wolfes type three-body problems whose 
exact solutions are given in terms of these EOPs. This is the task that we
have addressed in this paper. 

In particular, in this paper , we have added to the list of exactly solvable
three-body problems in one dimension by constructing a number of rationally extended three-body 
problems whose
eigenfunctions are in terms of exceptional Laguerre as well as Jacobi
polynomials (EOPs). Remarkably, we have also been able to construct complex 
rationally extended $PT$-invariant three-body problems whose solutions are  
not in the exact forms of EOPs rather they are written in the forms 
of some new type of polynomials which can be further expanded in terms of 
classical Jacobi polynomials.  
    
The plan of the paper is as follows:
In section $2$, we briefly recall the Calogero type as well as Wolfes type
three body problems and how the choice of Jacobi coordinates play a crucial
role in obtaining the exact solution of such three-body problems. 
In section $3$, we explain our ideas by extending the Calogero-Wolfes type 
three body problems by adding new three-body interaction terms 
and obtain their solution in terms of the product of the $X_1$ exceptional 
Laguerre and  $X_1$ exceptional Jacobi polynomials. The generalization to the 
$X_m$ and $X_p$ polynomials in Laguerre and Jacobi variables respectively is 
also discussed. In subsection $3.1$, we discuss another example where we add
 complex but $PT$-invariant three-body interaction terms and show that
the solution in general is product of $X_m$ Laguerre polynomials times some
new polynomial which can be expanded in terms of classical Jacobi orthogonal  
polynomials. 

A list of all possible rationally extended real and $PT$ symmetric 
complex three body problems whose solution is in terms of the product of the
$X_1$ Laguerre and $X_1$ Jacobi polynomials is given in  Table $1$.  
A similar list of all possible rationally extended real and $PT$ symmetric 
complex three body problems whose solution is in terms of the product of the
$X_m$ Laguerre and $X_p$ Jacobi polynomials is given in  Table $2$.  
Finally, we summarize our results in section $4$.

\section{The Calogero-Wolfes type three body problems}

The Calogero's \cite {cal_69}  exactly solvable three body problem is 
characterized by the potential
\be\label{cal}
V_{C}=V_{H}+V_{I},
\ee
where $V_H$ and $V_I$ are harmonic and inverse square potentials
\be\label{1}
V_{H}=\frac{\omega^2}{8}\sum_{i<j}(x_i-x_j)^2; \quad V_{I}=g\sum_{i<j}(x_i-x_j)^{-2},
\ee
respectively. Here $g>-1/2$ is a coupling parameter used to avoid a collapse of the system. This three body problem 
is solved exactly by defining the Jacobi co-ordinates
\be\label{jc_1}
R=\frac{1}{3}(x_1+x_2+x_3),\nonumber 
\ee 
and 
\be\label{jc_2}
x=\frac{(x_1-x_2)}{\sqrt{2}},\qquad y=\frac{(x_1+x_2-2x_3)}{\sqrt{6}}.
\ee
In polar co-ordinates
\be\label{jc_3}
x=r\sin\phi, \qquad y=r\cos\phi; \qquad 0\le r \le\infty, \quad 0\le\phi\le 2\pi,
\ee
with 
\be
r^2=\frac{1}{3}[(x_1-x_2)^2+(x_2-x_3)^2+(x_3-x_1)^2]; \qquad x_1\neq x_2 \neq x_3.
\ee
Using Eqs. (\ref{jc_2}) and (\ref{jc_3}), one can easily show that
\ba\label{jc_4}
&(x_1-x_2)=\sqrt{2}r\sin\phi,\nonumber\\
&(x_2-x_3)=\sqrt{2}r\sin(\phi+2\pi/3),\nonumber\\
&(x_3-x_1)=\sqrt{2}r\sin(\phi+4\pi/3).
\ea
In $1974$ Wolfes \cite{wolf_74} showed  that a three-body potential 
\be\label{vwg}
V_W(g)= g[(x_1+x_2-2x_3)^{-2}+(x_2+x_3-2x_1)^{-2}+(x_3+x_1-2x_2)^{-2}]
\ee
is also solvable when it is added to $V_C$ with or without the inverse square 
potential $V_{I}$. Note that the last two terms in the potential (\ref{vwg}) are 
 the cyclic permutation of the first one. Hence from now onwards, we will refer
to such terms as c.p. i.e. (cyclic permutation terms). 

\section{Rationally extended three body potentials} 

In this section, we follow the method adopted in \cite{kh_bh_93} and obtain 
some new three body potentials 
whose solutions are in terms of EOPs. As an illustration, we first consider 
a potential of the form
\be\label{pot_first}
V=V_{H}+V_{W}(g)+V_{int}+V_{rat},
\ee
where $V_H$ and $V_W(g)$ are as given by Eqs. (\ref{1}) and (\ref{vwg})
respectively 
while $V_{int}$ is given by 
\be\label{int}
V_{int}=\frac{3f_1}{2\sqrt{2}r}\bigg[\frac{(x_1-x_2)}{(x_1+x_2-2x_3)^2}+\mbox{c.p}\bigg].
\ee
Finally, the newly added rational term $V_{rat}$ is defined as   
\be
V_{rat}=V^{(1)}_{rat}+V^{(2)}_{rat},
\ee
where $V^{(1)}_{rat}$ ,  $V^{(2)}_{rat}$ are the two new rational
interaction terms introduced by us.
Thus the combined potential becomes 
\ba\label{pot_1}
V&=&\frac{\omega^2}{8}\sum_{i<j}(x_i-x_j)^2+3g[(x_1+x_2-2x_3)^{-2}+\mbox{c.p}]\nonumber\\
&+&\frac{3f_1}{2\sqrt{2}r}\bigg[\frac{(x_1-x_2)}{(x_1+x_2-2x_3)^2}+\mbox{c.p}\bigg] +V^{(1)}_{rat}+V^{(2)}_{rat}.
\ea     
Now we define $V^{(1)}_{rat}$ and $V^{(2)}_{rat}$ as
\be\label{pot_new_1}
V^{(1)}_{rat}=\frac{a\sum_{i<j}(x_i-x_j)^2+c_1}{\big(b\sum_{i<j}(x_i-x_j)^2+c_2\big)^2}
\ee
and 
\be\label{pot_new_2}
V^{(2)}_{rat}=\frac{\delta}{r^2}\bigg[\frac{k_1}{\big(k_2+k_3\xi\big)}-\frac{k_4}{\big(k_2+k_3\xi\big)^2}\bigg].
\ee
Here $a$, $b$, $c_1$, $c_2$, $\delta$, $k_1$, $k_2$, $k_3$, $k_4$ are 
constants while $\xi$ is a function of $x_1, x_2, x_3$. The values of 
these constants and the form of the function $\xi$ depend on the nature of 
the problems. For example, if the function $\xi$ is  of the forms  
\be\label{x_1}
\xi=\frac{4}{2\sqrt{2}r^3}(x_1-x_2)(x_2-x_3)(x_3-x_1)
\ee
or
\be\label{x_2}
\xi=-\frac{3}{\sqrt{2}r}\bigg(\sum_{i<j}(x_i-x_j)^{-1}\bigg)^{-1},
\ee
it leads to two different three body problems. On using the Jacobi 
co-ordinates (\ref{jc_1})-(\ref{jc_4}) and  the 
following identities 
\ba\label{id_1}
\prod^{3}_{s=1}\sin(\phi+\frac{2(s-1)\pi}{3})&=&-\frac{1}{4}\sin(3\phi)\,, 
\nonumber\\
\sum^3_{s=1}\cosec(\phi+\frac{2(s-1)\pi}{3})&=&3\cosec (3\phi)\,,
\ea 
remarkably, both the forms of the above $\xi$,  reduce to a same function
\be
\xi=-\sin(3\phi)\,.
\ee
In polar co-ordinates $(r,\phi)$, the Schr\"odinger equation corresponding 
to the above potential is given by $(\hbar=2m=1)$ 
\be\label{sch_1}
\bigg[-\frac{d^2}{dr^2}-\frac{1}{r}\frac{d}{dr}-\frac{1}{r^2}
\frac{d^2}{d\phi^2}\bigg]\psi_{n\ell}(r,\phi)+V\psi_{n\ell}(r,\phi)= E_{n\ell}\psi_{n\ell}(r,\phi)\,.
\ee
Note that the potential is non-central but separable and hence one can write
the wave function in the form
\be\label{wf_1}
\psi_{n\ell}(r,\phi)=\frac{R_{n\ell}(r)}{r^{1/2}}\Phi_{\ell}(\phi )\,.
\ee 
It is then straightforward to show that the radial component of the wave 
function satisfies the equation
\be\label{rad_1} 
\bigg[-\frac{d^2}{dr^2}+V(r)\bigg]R_{n\ell}(r)=E_{n\ell}R_{n\ell}(r),
\ee 
where the potential $V(r)$ is given by
\be\label{vr_1}
V(r)=V_{Con}(r)+V^{(1)}_{rat}(r)\,.
\ee
Here $V_{Con}(r)$ is the conventional radial oscillator potential, i.e.  
\be
V_{Con}(r)=\frac{3}{8}\omega^2 r^2 + \frac{(\lambda ^2_{\ell}-1/4)}{r^2}\,,
\ee
while the rational term  $V^{(1)}_{rat}(r)$ is given by
\be\label{rat_1}
V^{(1)}_{rat}(r) =\frac{(3ar^2+c_1)}{(3br^2+c_2)^2}\,.
\ee
Now notice that in case we set the parameters $a$, $b$, $c_1$ and $c_2$ as 
\be
a=2\omega^2;\qquad 3b=(\sqrt{3/2})\omega; \qquad c_1=-4(\sqrt{3/2})\omega c_2 
\qquad \mbox{and} \qquad c_2=2\lambda _{\ell}\,,
\ee
then the above potential (\ref{vr_1}) matches exactly with the rationally 
extended radial oscillator potential given in \cite {que}. As shown there, 
in that case, the bound states eigenfunctions are given by the $X_1$ Laguerre 
polynomials 
\be\label{wf_rad}
R_{n\ell}(r)\propto \frac{r^{\lambda _{\ell}+1/2}\exp\big(-\frac{1}{4}(\sqrt{3/2}\big)\omega r^2)}{\big[(\sqrt{3/2})\omega r^2+2\lambda _{\ell}\big]}\hat{L}^{(\lambda _\ell)}_{n+1}\big((\sqrt{3/8})\omega r^2\big),
\ee
and the corresponding energy eigenvalues are  
\be\label{en_1}
E_{n\ell}=(\sqrt{3/2})\omega (2n+\lambda _\ell +1 ); \qquad n=0,1,2,...; \qquad \ell=0,1,2...,
\ee
with $\lambda _\ell>0$. 

On the other hand, the angular part of the 
eigenfunction $\Phi_\ell(\phi)$ can be shown to satisfy the Schr\"odinger 
equation
\be\label{sch_phi}
\bigg[-\frac{d^2}{d\phi^2}+V(\phi ) \bigg ]\Phi_{\ell}(\phi )=\lambda ^2_\ell \Phi_{\ell}(\phi),
\ee 
where
\ba\label{pot_phi}
V(\phi )&=&\frac{g}{2}\sum^{3}_{s=1}\sec^2(\phi+\frac{2(s-1)\pi}{3})+\frac{f_1}{2}\sum^{3}_{s=1}\sec(\phi+\frac{2(s-1)\pi}{3})\nonumber\\
&\times &\tan(\phi+\frac{2(s-1)\pi}{3})+V^{(2)}_{rat}(\phi )\,.
\ea
Using the following identities
\be\label{id_2}
\sum^{3}_{s=1}\sec^2(\phi+\frac{2(s-1)\pi}{3})=9\sec^2(3\phi)\,,
\ee
and 
\be\label{id_3}
\sum^{3}_{s=1}\sec(\phi+\frac{2(s-1)\pi}{3})\tan(\phi+\frac{2(s-1)\pi}{3})
=-9\tan(3\phi)\sec(3\phi)\,,
\ee
the potential (\ref{pot_phi}) takes the form 
\be\label{pot_phi_2}
V(\phi)=V_{Con}(\phi)+V^{(2)}_{rat}(\phi)\,,
\ee
where $V_{Con}(\phi)$ is essentially the conventional trigonometric Scarf 
potential 
\be
V_{Con}(\phi)=\frac{9g}{2}\sec^2(3\phi)-\frac{9f_1}{2}\sec(3\phi)\tan(3\phi),
\ee
while the rational term $V^{(2)}_{rat}(\phi)$ takes the form 
\be
V^{(2)}_{rat}(\phi)=\delta\bigg[\frac{k_1}{\big(k_2-k_3\sin(3\phi)\big)}-\frac{k_4}{\big(k_2-k_3\sin(3\phi)\big)^2}\bigg].
\ee
On comparing the above obtained potential (\ref{pot_phi_2}) with the 
rationally extended trigonometric Scarf potential as given in \cite{que}, i.e. 
\ba\label{tsc_rat}
V(\phi) &=&[A(A-3)+B^2]\sec^2(3\phi)-B(2A-3)\sec(3 \phi)\tan(3 \phi)\nonumber\\
&+& 9\bigg[\frac{2(2A-3)}{(2A-3-2B\sin(3 \phi))}-\frac{2[(2A-3)^2-4B^2]}{(2A-3-2B\sin(3 \phi))^2}\bigg],
\ea
defined over the range $-\frac{\pi}{6}<\phi<\frac{\pi}{6}$ and the parametric 
restriction $0<B<A-3$, we get
 \ba\label{cond_1}
 &&A(A-3)+B^2=\frac{9g}{2};\nonumber\\
 &&B(2A-3)=\frac{9f_1}{2},
 \ea
 \be
 \delta=9;\quad k_1=2k_2;\quad k_2=2A-3; \quad k_3=2B; \quad  \mbox{and} \quad k_4=2(k^2_2-k^2_3).
 \ee
To get explicit expressions for $A$ and $B$ in terms of $g$ and $f_1$, we 
solve Eq. (\ref{cond_1}) and 
obtain four possible roots for $A$ and $B$. Out of all these, we consider one 
suitable root given by
\ba\label{2}
&&A=\frac{1}{16}[24+12\sqrt{2(1+2g+\zeta )} ]; \quad B=\frac{3f_1}{\sqrt{2(1+2g+\zeta )}};\nonumber\\
&&\mbox{where}\quad \zeta =\sqrt{(1+2g)^2-4f^2_1}.
\ea 
The bound state wave functions of $V(\phi)$ are well known and given in terms 
of $X_1$ exceptional Jacobi polynomial 
$\hat{L}^{(\alpha,\beta)}_{\ell+1}(\sin(3\phi))$ as 
\be\label{tsc_wf}
\Phi_{\ell}(\phi)\propto \frac{(1-\sin(3\phi))^{\frac{1}{6}(A-B)}(1+\sin(3\phi))^{\frac{1}{6}(A+B)}}{(2A-3-2B\sin(3\phi) )}\hat{P}^{(\alpha,\beta)}_{\ell+1}(\sin(3\phi)),
\ee
with the parameters 
\ba\label{parasc_1}
\alpha&=&\bigg(\frac{A}{3}-\frac{B}{3}-\frac{1}{2}\bigg)=\frac{1}{2}\sqrt{1+2(g-f_1)};\nonumber\\
 \beta&=&\bigg(\frac{A}{3}+\frac{B}{3}-\frac{1}{2}\bigg)=\frac{1}{2}\sqrt{1+2(g+f_1)},
\ea 
and the energy eigenvalues
\be\label{en_scarf}
\lambda ^2_{\ell}=(A+3\ell)^2;\quad \ell=0,1,2,...,.
\ee

Summarizing, for the three-body problem with the potentials as given by 
Eq. (\ref{pot_first}), the exact energy eigenvalues are given by 
Eq. (\ref{en_1}) where
using the expression for $A$ as given by Eq. (\ref{2}), $\lambda_l$ 
as given by Eq. (\ref{en_scarf}) takes the form
\be\label{3}
\lambda_\ell = \frac{3}{4}[2(2\ell+1)+\sqrt{2(1+2g+\zeta)}]\,,
\ee
with $\zeta$ being given by Eq. (\ref{2}). Note that $\zeta$ and hence 
$\lambda_\ell$
are real only if $f_1 < g+1/2$. The corresponding eigenfunctions are given by
Eq. (\ref{wf_1}) where $R_\ell(r)$ and $\Phi_\ell(\phi)$ are as given by 
Eqs. (\ref{wf_rad}) and (\ref{tsc_wf}) respectively. 

{\bf Generalization to the $X_m$ case:}

The above three body problems can be now easily extended to a more general 
three-body problem where the solutions of both the radial and the angular
parts is in terms of $X_m$ Laguerre and $X_p$ Jacobi polynomials by redefining 
Eqs. (\ref{pot_new_1}) and (\ref{pot_new_2}) as
\ba\label{vnew_1m}
V^{(1)}_{rat}\Rightarrow V^{(1)}_{rat}(m)&=&-2m(\sqrt{3/2})\omega-\frac{\omega^2}{2}\sum_{i<j}(x_i-x_j)^2\frac{L^{(\lambda _\ell+1)}_{m-2}(-(\sqrt{3/8})\omega r^2)}{L^{(\lambda _\ell-1 )}_{m}(-(\sqrt{3/8})\omega r^2)}\nonumber\\
&+&(\sqrt{3/2})\omega \bigg[(\sqrt{1/6})\omega \sum_{i<j}(x_i-x_j)^2+2\lambda _\ell-2\bigg] \frac{L^{(\lambda _\ell)}_{m-1}(-(\sqrt{3/8})\omega r^2)}{L^{(\lambda _\ell-1 )}_{m}(-(\sqrt{3/8})\omega r^2)}\nonumber\\
&+&\omega^2\sum_{i<j}(x_i-x_j)^2\bigg(\frac{L^{(\lambda _\ell)}_{m-1}(-(\sqrt{3/8})\omega r^2)}{L^{(\lambda _\ell-1 )}_{m}(-(\sqrt{3/8})\omega r^2)}\bigg)^2,
\ea
and
\ba\label{vnew_2m}
V^{(2)}_{rat}\Rightarrow V^{(2)}_{rat}(p)&=&\frac{\delta}{r^2}\bigg[-2p(\alpha-\beta-p+1)-(\alpha-\beta-p+1)(\alpha+\beta+(\alpha-\beta+1)z)\nonumber\\
&\times & \frac{P^{(-\alpha,\beta)}_{p-1}(z)}{P^{(-\alpha-1,\beta-1)}_{p}(z)}+\frac{(\alpha-\beta-p+1)^2\chi}{2}\bigg(\frac{P^{(-\alpha,\beta)}_{p-1}(z)}{P^{(-\alpha-1,\beta-1)}_{p}(z)}\bigg)^2\bigg],
\ea
where 
\be
z=-\xi \quad \mbox {and} \quad \chi=1-z^2,
\ee
with a constant $\delta=9$.
Similar to the $X_1$ cases, using the concept of Jacobi co-ordinates, 
we get the radial and angular dependent potentials 
\ba\label{vnew_1m2}
V^{(1)}_{rat}(r)\Rightarrow V^{(1)}_{rat}(m,r)&=&-2m(\sqrt{3/2})\omega-\frac{3\omega^2 r^2}{2}\frac{L^{(\lambda _\ell+1)}_{m-2}(-(\sqrt{3/8})\omega r^2)}{L^{(\lambda _\ell-1 )}_{m}(-(\sqrt{3/8})\omega r^2)}\nonumber\\
&+&(\sqrt{3/2})\omega \bigg[(\sqrt{3/2})\omega r^2 +2\lambda _\ell-2\bigg] \frac{L^{(\lambda _\ell)}_{m-1}(-(\sqrt{3/8})\omega r^2)}{L^{(\lambda _\ell-1 )}_{m}(-(\sqrt{3/8})\omega r^2)}\nonumber\\
&+&3\omega^2 r^2\bigg(\frac{L^{(\lambda _\ell)}_{m-1}(-(\sqrt{3/8})\omega r^2)}{L^{(\lambda _\ell-1 )}_{m}(-(\sqrt{3/8})\omega r^2)}\bigg)^2,
\ea
and
\ba\label{vnew_2m2}
V^{(2)}_{rat}(\phi)\Rightarrow V^{(2)}_{rat}(p,\phi )&=&9 \bigg[-2p(\alpha-\beta-p+1)-(\alpha-\beta-p+1)\nonumber\\
&\times &(\alpha+\beta+(\alpha-\beta+1)\sin(3\phi))\frac{P^{(-\alpha,\beta)}_{p-1}(\sin(3\phi) )}{P^{(-\alpha-1,\beta-1)}_{p}(\sin(3\phi))}\nonumber\\
&+& \frac{(\alpha-\beta-p+1)^2\cos^2(3\phi)}{2}\bigg(\frac{P^{(-\alpha,\beta)}_{p-1}(\sin(3\phi) )}{P^{(-\alpha-1,\beta-1)}_{p}(\sin(3\phi))}\bigg)^2\bigg],\nonumber\\
\ea
respectively. The solutions of the Schr\"odinger equation for these potentials $V(r)\Rightarrow V(r,p)$ and 
$V(\phi)\Rightarrow V(p,\phi)$ in terms of $X_m$ Laguerre and $X_p$ Jacobi 
EOPs respectively are given as 
\ba\label{rad_psi_m}
R_{n\ell}(m,r)\propto \frac{r^{\lambda _{\ell}+1/2}\exp\big(-\frac{1}{4}(\sqrt{3/2}\big)\omega r^2)}{L^{(\lambda _\ell-1)}_{m}(-(\sqrt{3/8})\omega r^2)}\hat{L}^{\lambda _\ell}_{n+m}\big((\sqrt{3/8})\omega r^2\big),
\ea   
and 
\be\label{tsc_psi_m}
\Phi_{\ell}(p,\phi)\propto \frac{(1-\sin(3\phi))^{\frac{1}{6}(A-B)}(1+\sin(3\phi))^{\frac{1}{6}(A+B)}}{P^{(-\alpha-1,\beta-1)}_{p}(\sin(3\phi))}\hat{P}^{(\alpha,\beta)}_{\ell+p}(\sin(3\phi)).
\ee
The energy eigenvalues will be same as given in Eqs. (\ref{en_1}), 
(\ref{en_scarf}) and Eq. (\ref{3}).. 

By redefining the constants $\delta$, $k_1$, $k_2$, $k_3$, $k_4$ and the 
function $\xi$ which is written in terms of $x_1, x_2$ and $x_3$, two more 
three-body exactly solvable rationally 
extended real problems with their corresponding mapping potentials
the RE trigonometric P\"oschl-Teller II and RE trigonometric P\"oschl-Teller 
potentials respectively 
are constructed whose solutions are in terms of the products of the exceptional
$X_1$ Laguerre and  $X_1$ Jacobi EOPs. 
The details are mentioned in  Table $1$. In the same way by redefining one 
more function i.e. $\chi$, these two real potentials 
are further generalized to the product of $X_m$ Laguerre and $X_p$ Jacobi 
EOPs. The deails of these are given in Table $2$. 

\subsection{Rationally extended three-body $PT$ symmetric complex potentials}

In this section, we discuss some interesting rationally extended three body 
complex potentials  whose mapping potentials
are not real but they are complex and $PT$ symmetric. The solutions of these 
potentials are not in 
the exact form of EOPs rather they are written in the form of some types of new polynomials discussed in Ref. \cite {cq_12} which are further  expanded in 
terms of classical Jacobi polynomials.

Let us consider a complex potential $V$ of the form 
\ba
V=V_H+V_I+V_{int}+V_{rat},
\ea
where $V_H$ and $V_I$ are given by Eq. (\ref{1}), while $V_{int}$ is given by
\be
V_{int}=\frac{\sqrt 3}{2r^2}if_1\bigg[\frac{(x_1+x_2-2x_3)}{(x_1-x_2)}+\mbox{c.p}\bigg].
\ee
Using $V_{rat}=V^{(1)}_{rat}+V^{(2)}_{rat}$, the combined potential becomes
\ba\label{potpt_1}
V&=&\frac{\omega^2}{8}\sum_{i<j}(x_i-x_j)^2+g\sum_{i<j}(x_i-x_j)^{-2}\nonumber\\
&+&\frac{\sqrt 3}{2r^2}if_1\bigg[\frac{(x_1+x_2-2x_3)}{(x_1-x_2)}+\mbox{c.p}\bigg] +V^{(1)}_{rat}+V^{(2)}_{rat}.
\ea    
Here $V^{(1)}_{rat}$ is again as given in Eq. (\ref{pot_new_1}), however the 
form of $V^{(2)}_{rat}$ depends on 
the nature of the function $\xi$. If we define 
\be
\xi=\frac{1}{3\sqrt 3}\bigg(\frac{x_1+x_2-2x_3}{x_1-x_2}+\mbox{c.p}\bigg),
\ee
and using the identities
\be
\sum^3_{s=1}\cosec^2(\phi+\frac{2(s-1)\pi}{3})=9\cosec^2(3\phi)
\ee
and 
\be
\sum^3_{s=1}\cot(\phi+\frac{2(s-1)\pi}{3})=3\cot(3\phi),
\ee
then the $\phi$-dependent potential will be 
\be\label{V2_phi}
V(\phi)=V_{Con}(\phi) + V^{(2)}_{rat}(\phi),
\ee
with the equivalent conventional $PT$ symmetric trigonometric Eckart potential
\be
V_{Con}(\phi)=\frac{9}{2}g\cosec^2(3\phi)+\frac{9}{2}if_1\cot(3\phi),
\ee
and the rational term
\be
V^{(2)}_{rat}(\phi)=\delta \bigg[\frac{k_1}{(k_2+k_3\cot(3\phi) )}
+\frac{k_4}{(k_2+k_3\cot(3\phi) )^2}\bigg].
\ee
This potential is equivalent to the rationally extended $PT$ symmetric 
complex trigonometric Eckart potential\footnote{Which is 
easily obtained by complex co-ordinate transformation $x\rightarrow ix$ of the 
rationally extended hyperbolic Eckart potential given in \cite{cq_12}.}
\ba\label{V2_tek_pt}
V(\phi)&=&A(A-3)\cosec^2(3\phi)+2iB\cot(3\phi)+\frac{9}{A^2(A-3)^2}\bigg[\frac{-4iB[A^2(A-3)^2-B^2]}{(iB+A(A-3)\cot(3\phi) )}\nonumber\\
&+& \frac{2[A^2(A-3)^2-B^2]^2}{(iB+A(A-3)\cot(3\phi) )^2}\bigg],
\ea
with the constants
\ba
\delta=9;\quad k_1&=&\frac{-4iB[A^2(A-3)^2-B^2]}{A^2(A-3)^2}; \quad k_2=iB; \quad k_3=A(A-3);\nonumber\\
\mbox{and} \quad k_4&=&\frac{2[A^2(A-3)^2-B^2]^2}{A^2(A-3)^2}.
\ea
The potential parameters $A$ and $B$ in terms of $g$ and $f_1$ are related as
\ba
A&=&\frac{3}{2}+3a; \quad B=\frac{9}{4}f_1,\nonumber\\
\mbox{where}\quad a&=&\frac{1}{2}\sqrt{1+2g}.
\ea
By $P$ (i.e. parity) we mean here $\phi \rightarrow \phi+\pi$ while by 
$T$ (i.e. time reversal) we mean $t \rightarrow -t$ and $i \rightarrow -i$. 
The wavefunction associated with the above potential is product of the
$X_1$ Laguerre polynomial as given by Eq. (\ref{wf_rad}) times $\Phi_{\ell}(\phi)$ 
which is given by
\be\label{wfpt_eck}
\Phi_\ell(\phi)\propto \frac{(z-1)^\frac{\alpha_\ell}{2}(z+1)^\frac{\beta_\ell}{2}}{(iB+A(A-3)\cot(3\phi) )}y^{(A/3,B/3)}_\ell(z),
\ee
with $z=i\xi=i\cot(3\phi)$. Here the polynomial function $y^{(A/3,B/3)}_\ell(z)$ 
can be expressed in terms of the classical Jacobi 
polynomials $P^{(\alpha_\ell,\beta_\ell)}_{\ell}(z)$ as
\ba
y^{(A/3,B/3)}_\ell(z)&=&\frac{2(\ell+\alpha_\ell)(\ell+\beta_\ell)}{(2\ell+\alpha_\ell+\beta_\ell)}q^{(A/3,B/3)}_1(z)P^{(\alpha_\ell,\beta_\ell)}_{\ell-1}(z)\nonumber\\
&-&\frac{2(1+\alpha_1)(1+\beta_1)}{(2+\alpha_1+\beta_1)}P^{(\alpha_\ell,\beta_\ell)}_{\ell}(z).
\ea
The parameters $\alpha_\ell$ and $\beta_\ell$ in terms of $A$ and $B$  are 
given by
\be
\alpha_\ell=-(A/3-1+\ell)+\frac{B/9}{(A/3-1+\ell)}; \quad \beta_\ell=-(A/3-1+\ell)-\frac{B/9}{(A/3-1+\ell)}.
\ee
and $q^{(A/3,B/3)}_1(z)=P^{(\alpha_1,\beta_1)}_{1}(z)$ (Classical Jacobi 
polynomial for $\ell=1$). 
The energy eigenvalues are given by Eq. (\ref{en_1})
where $\lambda_l$ is given by 
\be\label{enpt_1}
\lambda ^2_\ell=9(\ell+a-\frac{1}{2})-\frac{9f^2_1}{16(\ell+a-\frac{1}{2})^2}; \quad \ell=0,1,2,...
\ee
Note $\Phi_\ell$ has to be in such order that $\lambda_{\ell}^{2} > 0, \quad f_1<4(\ell+a-1/2)$.

Similar to the EOPs cases, the above complex potential can be generalized  
for any non-zero positive values of $p$ by again considering the radial 
potential $V^{(1)}_{rat}(m)$ as given by Eq. (\ref{vnew_1m2}) and 
defining $V^{(2)}_{rat}(p)$ by 
\ba
V^{(2)}_{rat}\Rightarrow V^{(2)}_{rat}(p)&=&\delta \sum_{i<j}(x_i-x_j)^{-2}\bigg[ 2i\xi \frac{\dot{q}^{(A/3,B/3)}_p(z)}{q^{(A/3,B/3)}_p(z)}-\frac{8r^2}{9}\sum_{i<j}(x_i-x_j)^{-2}\nonumber\\
&\times & \Bigg( \frac{\ddot{q}^{(A/3,B/3)}_p(z)}{q^{(A/3,B/3)}_p(z)}-\bigg( \frac{\dot{q}^{(A/3,B/3)}_p(z)}{q^{(A/3,B/3)}_p(z)}\bigg )^2 \Bigg )-p\bigg ].
\ea
For $\delta=16$ and using Jacobi co-ordinates, we get 
 \ba
V^{(2)}_{rat}\Rightarrow V^{(2)}_{rat}(p,\phi)&=& -18 \cosec^2(3\phi)\bigg[ 2i\cot(3\phi) \frac{\dot{q}^{(A/3,B/3)}_p(z)}{q^{(A/3,B/3)}_p(z)}-\cosec^2(3\phi)\nonumber\\
&\times & \Bigg( \frac{\ddot{q}^{(A/3,B/3)}_p(z)}{q^{(A/3,B/3)}_p(z)}-\bigg( \frac{\dot{q}^{(A/3,B/3)}_p(z)}{q^{(A/3,B/3)}_p(z)}\bigg )^2 \Bigg )-p\bigg ].
\ea 
The wavefunctions associated with this potentials will be product of $X_m$
Laguerre times $\Phi_{\ell}(p,\phi)$ which is given by  
\be\label{wfpt_eckm}
\Phi_\ell(p,\phi)\propto \frac{(z-1)^\frac{\alpha_\ell}{2}(z+1)^\frac{\beta_\ell}{2}}{q^{(A/3,B/3)}_m(z)}y^{(A/3,B/3)}_{\nu,p}(z); \quad \nu=\ell+p-1,
\ee
where $q^{(A/3,B/3)}_p(z)=P^{(\alpha_p,\beta_p)}_p$ and the polynomial 
function $y^{(A/3,B/3)}_{\nu,p}(z)$ is 
\ba
y^{(A/3,B/3)}_{\nu,p}(z)&=&\frac{2(\ell+\alpha_\ell)(\ell+\beta_\ell)}{(2\ell+\alpha_\ell+\beta_\ell)}q^{(A/3,B/3)}_p(z)P^{(\alpha_\ell,\beta_\ell)}_{\ell-1}(z)\nonumber\\
&-&\frac{2(p+\alpha_p)(p+\beta_p)}{(2p+\alpha_p+\beta_p)}q^{(A/3+1,B/3)}_{p-1}(z)P^{(\alpha_\ell,\beta_\ell)}_{\ell}(z),
\ea
with the parameters
 \be
\alpha_p=-(A/3-1+p)+\frac{B/9}{(A/3-1+p)}; \quad  \beta_p =-(A/3-1+p)-\frac{B/9}{(A/3-1+p)}.
\ee
The energy eigenvalues are again given by Eq. (\ref{en_1})
where $\lambda_l$ is given by Eq. (\ref{enpt_1}).

Similar to the real cases, another $3$-body complex problem with the  $PT$ 
symmetric complex mapping potential
i.e the RE $PT$ symmetric  complex trigonometric Rosen-Morse potential can
 also be constructed by redefining all constant parameters 
and the function $\xi$. The details about this potential 
for $p=1$ and then for any arbitrary values of $p$ along with the other 
relevant parameters are given in detail in tables $1$ and $2$ 
respectively.

\begin{sidewaystable}[!htb]\small\centering
\begin{tabular}{|l|*4{c|}} \hline
\pbox{30cm} {\textbf{Forms of potential } \\
\textbf{(V)} }&\pbox{30cm}{ \textbf{Function $\xi $ and other } \\
\textbf { constant parameters} }& $V(\phi )=V_{Con} (\phi ) +V^{(2)}_{rat}(\phi )$ &\pbox{20cm} { $\lambda _{\ell}^{2}$ \\} 
&\pbox{20cm} { $\Phi_{\ell}(\phi )$\\ } \\ \hline
\pbox{10cm} {(I) $V = V_{H} + V_{W}(g)+ V_{int}$\\
$+ V_{rat}$, \\
 $V_{W}(g)= $ \\
$ 3g[(x_{1}+x_{2}-2x_{3})^{-2}$ +c.p],\\
$V_{int} = f_{1}\sum(x_{i}-x_{j})^{-2}$.}
&\pbox{10cm} {$ \xi = \frac{9}{6r^{2}}[(x_{1}+x_{2}-2x_{3})^{-1} $\\
$+c.t]^{-2}$\\
or $\xi =(\frac{4}{6\sqrt{6}})^{2}\frac{1}{r^{6}}[(x_{1}+x_{2}-2x_{3})$\\
$\times (x_{2}+x_{3}-2x_{1})(x_{1}+x_{3}-2x_{2})]^{2}$, \\
$\delta=36; k_{1}=2(\beta +\alpha );$\\
$k_{2}=2\beta; k_{3}=-2(\beta -\alpha )$\\
$k_{4}=2[(\beta +\alpha )^{2}-(\beta -\alpha )^{2}]$ }&
\pbox{10cm} {$V_{Con}(\phi )=\frac{9g}{2}\sec^{2}(3\phi ) $ \\
$+\frac{9f_1}{2} cosec^{2}(3\phi )$; \\
$V^{(2)}_{rat}(\phi )=36[\frac{2(\beta +\alpha )}{(2\beta -2(\beta -\alpha )\cos^{2}(3\phi ))}$ \\
$-\frac{2[(\beta +\alpha )^{2}-(\beta -\alpha )^{2}]}{(2\beta -2(\beta -\alpha )\cos^{2}(3\phi ))^{2}}]$;\\
 $\alpha = \frac{1}{2}\sqrt {1+2f_{1}},$ \\
$ \beta = \frac{1}{2}\sqrt {1+2g}, $ \\
($V(\phi )$ =RE Trigonometric  \\
P\"oschl- Teller $II$ potential)} & \pbox{20cm} {$9(\alpha +\beta +1+2\ell)^2$ \\} & \pbox{20cm} { $\propto \frac{(1- z )
^{\frac{\alpha }{2} + \frac{1}{4}}(1 +z )^{\frac{\beta  }{2} + \frac{1}{4}}}
{(2\beta-2(\beta -\alpha )\xi)}$\\
$\times \hat{P}_{\ell+1}^{(\alpha ,\beta) }(z)  $; \\ 
$z=(2\xi-1)=\cos(6\phi)$.} \\ \hline
\pbox{10cm} {$(II) V = V_{C} + V_{int} + V_{rat},$ \\
$V_{int} = \frac{-f_{1}}{(\sqrt 6)r}[\frac{(x_{1}+x_{2}-2x_{3})}{(x_{1}-x_{2})^2} $ \\
$ +c.p].$}
&\pbox{10cm} {$ \xi = (\frac{3}{\sqrt{6}})\frac{1}{r}[(x_{1}+x_{2}-2x_{3})^{-1}] $ \\
$+c.t]^{-1}$\\
or $\xi =-\frac{4}{(6\sqrt{6})r ^{3}}(x_{1}+x_{2}-2x_{3})$\\
$\times(x_{2}+x_{3}-2x_{1})(x_{1}+x_{3}-2x_{2});$ \\
$\delta=9; k_{1}=2(\beta +\alpha );$\\
$k_{2}=(\beta +\alpha) ;k_{3}=(\beta -\alpha );$\\
$k_{4}=2[(\beta +\alpha )^{2}-(\beta -\alpha )^{2}].$}&
\pbox{10cm} {$V_{Con}(\phi ) = \frac{9g}{2} cosec^{2}(3\phi ) $ \\
$-\frac{9f_1}{2} cosec(3\phi )cot (3\phi );$ \\
$V^{(2)}_{rat}(\phi)=9[\frac{2(\beta +\alpha )}{(\beta+\alpha ) -(\beta -\alpha )\cos(3\phi)}$ \\
$-\frac{2[(\beta +\alpha )^{2}-(\beta -\alpha )^{2}]}{[(\beta+\alpha ) -(\beta -\alpha )\cos(3\phi )]^2},$\\
$\alpha = [\frac{1}{2}(g+f_{1}+\frac{1}{2})]^{\frac{1}{2}}, $ \\
$ \beta = [\frac{1}{2}(g-f_{1}+\frac{1}{2})]^{\frac{1}{2}}; f_1<g+\frac{1}{2}. $ \\
($V(\phi )$ = RE trigonometric \\
P\"oschl - Teller potential)} &\pbox{20cm} {$9(\frac{\alpha +\beta +1}{2}+\ell)^2$\\} & \pbox{20cm} 
{$\propto \frac{(1- z )^{\frac{\alpha }{2} + \frac{1}{4}}(1 +z )^{\frac{\beta  }{2} + \frac{1}{4}}}
{((\beta+\alpha )-(\beta -\alpha )z)} $ \\
$\times\hat{P}_{\ell+1}^{(\alpha ,\beta) }(z)  $, \\
$z=-\xi=\cos(3\phi)$ } \\ \hline
\pbox{10cm} {$(III) V = V_{H} + V_{W}(g)+ $\\
$V_{int}+ V_{rat},$ \\
$V_{int} = \frac{-3 \sqrt {3}}{2r^{2}}if_1[\frac{(x_{1}-x_{2})}{(x_{1}+x_{2} -2x_{3})} $ \\
$ +c.p].$}
&\pbox{10cm} {$\xi =\frac{1}{\sqrt{3}}( \frac{x_{1}-x_{2}}{x_{1}+x_{2}-2x_{3}}+ c.p)$\\
$k_{1}= \frac{4B}{A^{2}(A+3)^2}[A^{2}(A+3)^{2}- B^{2}] ; $ \\
$ k_{2}= B; k_{3}= i A (A + 3);$\\
$k_{4}=\frac{2}{A^{2}(A+3)^2}[A^{2}(A+3)^{2}- B^{2}]^{2};$ \\
$\delta = -9. $ }&
\pbox{10cm} {$V_{Con}(\phi ) = A(A+B) \sec ^2(3\phi) $ \\
$-2iB \tan (3\phi);  $ \\
$V^{(2)}_{rat}(\phi)= -\frac{9}{A^{2}(A+3^{2})} $ \\
$\times [\frac{4B(A^{2}(A+3)^{2}-B^{2})}{(B+iA(A+3)\tan(3\phi ))} + $ \\
$ \frac{2(A^{2}(A+3)^{2}-B^{2})^{2}}{(B+iA(A+3)\tan (3\phi ))^{2}}];$ \\
$A = \frac{3}{2}(\sqrt {1+2g}-1); g>-\frac{1}{2},$ \\
$B = \frac{9f_{1}}{4}.$ \\
($(V(\phi)=$ RE $PT$ symmetric  \\
 trigonomtric RM potential)} & \pbox{20cm} {$(A+3-3\ell)^{2}$ \\
$+ \frac{B^{2}}{(A+3-3\ell)^{2}}$} & \pbox{30cm}{
$\propto \frac{(1-z)^{\frac{\alpha _{\ell}}{2}}(1+z)^\frac{\beta _{\ell}}{2}}{(B +iA(A+3)\xi)}y_{\nu }^{(A/3,B/3)}(z)$;  \\ 
$ y_{\nu }^{(A/3,B/3)}(z) = \frac {2(\ell+\alpha _{\ell})(\ell+\beta _{\ell})}{2\ell+\alpha _{\ell}+\beta _{\ell}}$  \\
$\times q_{1} ^{(A/3,B/3)}(z) P_{\ell-1}^{(\alpha _{\ell},\beta _{\ell})}(z) - $  \\ 
$\frac {2(1+\alpha _{1})(1+\beta _{1})}{(2+\alpha _{1}+\beta _{1})} P_{\ell}^{(\alpha _{\ell},\beta _{\ell})}(z); $ \\
$\alpha _{\ell} =\frac{A}{3}+1-\ell +\frac {{B/9}}{\frac{A}{3}+1-\ell};$ \\
$\beta  _{\ell} =\frac{A}{3}+1-\ell - \frac {{B/9}}{\frac{A}{3}+1-\ell};$ \\ 
 $q_{1} ^{(A/3,B/3)} (z) = P_{1} ^ {(\alpha _{1}, \beta _{1})}(z ); $\\
$z =i\xi=i\tan (3\phi).$ }\\ \hline

\end{tabular}
\caption{Rationally extended (RE) three body problems and their equivalent RE real and $PT$ symmetric complex potentials. The $\phi$ dependent eigenfunctions 
($\Phi_{\ell}(\phi)$) of these potentials  are written in terms of $X_1$ EOPs ($\hat{P}_{\ell+1}^{(\alpha ,\beta) }(z)$) or in terms of 
some other type of polynomials $y_{\nu }^{(A/3,B/3)}(z)$. The terms $V_C$ and $V_H$ are same as given by Eqs. (\ref{cal}) and (\ref{1}) in the text. 
The complete  wave functions and the corresponding energy eigenvalues are given by Eqs. (\ref{wf_1}) and (\ref{en_1}). }
\end{sidewaystable}

\begin{sidewaystable}[!htb]\small\centering
\begin{tabular}{|l|*4{c|}} \hline
\pbox{20cm} {\textbf{Forms of potential} \\
\textbf{$(V(m))$} }&\pbox{20cm}{ \textbf{Forms of $\chi $ or $z$} } & \pbox {20cm}{$V(p,\phi) =V_{Con} (\phi ) +V^{(2)}_{rat}(p,\phi )$ } &\pbox{20cm} {  $\Phi_{\ell,p}(\phi )$\\ } \\ \hline
\pbox{10cm} {(I) $V(p) = V_{H} + V_{W}(g)+ V_{int}$\\
$+ V_{rat}(p) $}
&\pbox{10cm} {$\chi=z^2-1;\quad z=2\xi-1$
}&
\pbox{10cm} {
$V^{(2)}_{rat}(p,\phi)=36\big[-2p(\alpha -\beta -p+1)$\\
$-(\alpha -\beta -p+1)(\alpha +\beta +(\alpha -\beta +1)z)$\\
$\times \frac{P_{p-1}^{(-\alpha ,\beta) }(z)}{P_{p}^{(-\alpha -1,\beta -1)}(z)}+\frac{(\alpha -\beta -p+1)^{2}\chi}{2} $ \\
$\times(\frac{P_{p-1}^{(-\alpha ,\beta) }(z)}{P_{p}^{(-\alpha -1,\beta -1)}(z)})^{2} \big]$ \\
( RE Trigonometric P\"oschl-Teller II \\
potential)}  & \pbox{20cm} { 
$\propto \frac{(1- z )^{\frac{\alpha }{2} + \frac{1}{4}}(1 +z )^{\frac{\beta  }{2} + \frac{1}{4}}}
{P_{p-1}^{(-\alpha -1,\beta -1)}(z)}\hat{P}_{\ell+p}^{(\alpha ,\beta) }(z).$ \\ } \\ \hline
\pbox{10cm} {(III) $V(p) = V_{C} + V_{int} + V_{rat}(p)$}
&\pbox{10cm} {$\chi=z^2-1;\quad z=-\xi$
}&
\pbox{10cm} {
$V^{(2)}_{rat}(p,\phi)=9[-2p(\alpha -\beta-p+1)$\\
$-(\alpha -\beta -p+1)(\alpha +\beta +(\alpha -\beta +1)z)$ \\
$\times \frac{P_{p-1}^{(-\alpha ,\beta) }(z)}{P_{p}^{(-\alpha -1,\beta -1)}(z)}+\frac{(\alpha -\beta -p+1)^{2}\chi}{2}$ \\
$\times(\frac{P_{p-1}^{(-\alpha ,\beta )}(z)}{P_{p}^{(-\alpha -1,\beta -1)}(z)})^{2}]$ \\
( RE Trigonometric P\"oschl-Teller\\
 potential)} & \pbox{20cm} {
$ \propto \frac{(1- z )^{\frac{\alpha }{2} + \frac{1}{4}}(1 +z )^{\frac{\beta  }{2} + \frac{1}{4}}}
{P_{p-1}^{(-\alpha -1,\beta -1)}(z)}\hat{P}_{\ell+p}^{(\alpha ,\beta) }(z))$ \\ } \\ \hline
\pbox{10cm} {(III)$ V(p)=V_{H} + V_{W}(g)+V_{int}$\\
$+ V_{rat}(p).$ \\
}
&\pbox{10cm} {$z=i\xi$
}&
\pbox{10cm} {
$V^{(2)}_{rat}(p,\phi)= -18(1+\xi^2)[2i\xi\frac{\dot{q}^{(A/3,B/3)}_p(z)}{q^{(A/3,B/3)}_p(z)}- $ \\
$ (1+\xi)^2\big( \frac{\ddot{q}^{(A/3,B/3)}_p(z)}{q^{(A/3,B/3)}_p(z)}-(\frac{\dot{q}^{(A/3,B/3)}_p(z)}{q^{(A/3,B/3)}_p(z)})^2\big)-p];$ \\
(RE $PT$ symmetric trigonomtric\\
 RM potential)}& \pbox{30cm}{
$\propto \frac{(1-z)^{\frac{\alpha _{\ell}}{2}}(1+z)^\frac{\beta _{\ell}}{2}}{q^{(A/3,B/3)}_p(z)} y_{\nu,p }^{(A/3,B/3)}(z)$;  \\ 
$ y_{\nu,p }^{(A/3,B/3)}(z) = \frac {2(\ell+\alpha _{\ell})(\ell+\beta _{\ell})}{2\ell+\alpha _{\ell}+\beta _{\ell}}q_{p} ^{(A/3,B/3)}(z)$\\
$\times P_{\ell-1}^{(\alpha _{\ell},\beta _{\ell})}(z) -\frac {2(p+\alpha _{p})(p+\beta _{p})}{(2p+\alpha _{p}+\beta _{p})}$\\ 
$\times q^{(A/3-1,B/3)}_{p-1}(z)P_{\ell}^{(\alpha _{\ell},\beta _{\ell})}(z). $ \\
 }\\ \hline

\end{tabular}
\caption{RE three body problems and their $\phi$ dependent eigenfunctions ($\Phi_{\ell}(\phi)$) for any values of $p$ are written in terms of $X_p$ EOPs ($\hat{P}_{\ell+p}^{(\alpha ,\beta) }(z)$) or in terms of some other type of polynomials $y_{\nu,p }^{(A,B)}(z)$. The $\lambda_\ell$ will be same as that of the $m=1$ case. The complete  wave functions and the corresponding energy eigenvalues are given by Eqs. (\ref{wf_1}) and (\ref{en_1}).}
\end{sidewaystable}


\section{Summary and discussion}
 In this paper we have constructed several exactly solvable, rationally 
extended, Calogero-Wolfes type three body problems by adding new types of 
interaction terms. In particular, we have considered three extended problems
and using the Jacobi co-ordinates, these extended problems have been 
transformed into the equivalent forms of RE exactly solvable real as well 
as $PT$ symmetric complex potentials.  We have constructed new potentials whose
solutions are given as the product of the $X_1$ Laguerre times $X_1$
Jacobi polynomials. We have further generalized these potentials such that
there solution is given in terms of the product of $X_m$ Laguerre times 
$X_p$ Jacobi polynomials ($m,p = 1,2,3,...$). While some of these potentials
are real, others are complex but $PT$-symmetric potentials.

{\bf Acknowledgments}
 B.P.M. acknowledges the financial support 
from the Department of Science and Technology (DST), Gov. of India under SERC project sanction
grant No. $SR/S2/HEP-0009/2012$. A.K. wishes to thank Indian National Science Academy (INSA) for 
the award of INSA senior scientist position at Savitribai Phule Pune University.


\begin{thebibliography}{99}
\bibitem{cal_69} F. Calogero, {\it J. Math. Phys.} \textbf {10} (1969) 2191.
\bibitem{cal_71} F. Calogero, {\it J. Math. Phys.} \textbf {12} (1971) 419.
\bibitem{suth_71} B. Sutherland, {\it J. Math. Phys.} \textbf {12} (1971) 246.

\bibitem{cal_mar_74} F. Calogero and C. Marchioro, {\it J. Math. Phys.} \textbf {15} (1974) 1425.
\bibitem{ols_pere} M. A. Olshanetsky and A. M. Perelomov, {\it Phys. Rep. } \textbf {71} (1981) 314; ibid {\bf 94} (1983) 6.
\bibitem{poly_92} A. P. Polychronakos, {\it Phys. Rev. Lett.} \textbf {69} (1992) 703.
\bibitem{shas_88} B. S. Shastry, {\it Phys. Rev. Lett. } \textbf {60} (1988) 639.
\bibitem{hald_88} F. D. M. Haldane, {\it Phys. Rev. Lett. } \textbf {60} (1988) 635.
\bibitem{frahm_93} H. Frahm, {\it J. Phys. A } \textbf {26} (1993) L473.
\bibitem{wolf_74} J. Wolfes {\it J. Math. Phys. } \textbf {15} (1974) 1420.
\bibitem{kh_bh_93} A. Khare and R. K. Bhaduri, {\it J. Phys. A } \textbf{27} (1994) 2213.
\bibitem{cks}  F. Cooper, A. Khare, U. Sukhatme {\it  Phys. Rep.  } \textbf{251} (1995) 267; {\it "SUSY in Quantum Mechanics"} World Scientific (2001).
\bibitem{bender} C. M. Bender, S. Boettcher, {\it Phys. Rev. Lett.} \textbf{80} (1998) 5243.\\
                 C. M. Bender, S. Boettcher, {\it J. Phys. A} \textbf{31} (1998) L273.
 \bibitem{bpm} B Basu-Mallick, T Bhattacharyya, B P Mandal,
Modern Physics Letters A 20 (07) (2005), 543.
 \bibitem{bpm1}  B Basu-Mallick,  B P Mandal, Phys. Lett. A 284 (2001) 231.
 
  \bibitem{bpm2}   B Basu-Mallick, T Bhattacharyya, A Kundu, B P Mandal ,
  Czech. J. Phys. 54 (2004) 5.
              
                 
\bibitem{mz_08} M. Znojil and M. Tater, {\it J. Phys. A: Math. Gen. } \textbf{34} (2001) 1793. 
\bibitem{dnr1} D. Gomez-Ullate, N. Kamran and R. Milson, {\it J. Math. Anal.Appl.} \textbf{359} (2009) 352.  
\bibitem{dnr2} D. Gomez-Ullate, N. Kamran and R. Milson, {\it J. Phys. A} \textbf{43} (2010) 434016. 
\bibitem{xm1} D. Gomez-Ullate, N. Kamran and R. Milson, {\it J. Phys. A}  43 (2010) 434016.  
\bibitem{xm2} D. Gomez-Ullate, N. Kamran and R. Milson, {\it Contemporary Mathematics} {\bf 563} 51 2012.
\bibitem{que}  C. Quesne, {\it J.Phys.A} \textbf{41} (2008) 392001.
\bibitem{bqr}  B. Bagchi, C. Quesne and R. Roychoudhary, 
{\it Pramana J. Phys.} \textbf{73}(2009) 337, C. Quesne, SIGMA {\bf 5} (2009)
84.
\bibitem{os}  S. Odake and R. Sasaki, {\it Phys. Lett. B}, \textbf{684} 
(2010) 173; ibid {\bf 679} (2009) 414. {\it J. Math. Phys}, \textbf{51}, 053513 (2010).
\bibitem{hos} C-L. Ho, S ODAKE and R Sasaki, {\it SIGMA} \textbf{7} (2011) 107. 
\bibitem{hs} C-L. Ho and R Sasaki, {\it ISRN Math. Phys.} 2012 (2012) 920475. 
\bibitem{gom} D. Gomez-Ullate, N. Kamran and R. Milson, {\it J. Math. Anal. Appl.} 399(2) (2013) 480.
\bibitem{qu} C. Quesne, {\it Int. J. Mod. Phys. A} \textbf{26} (2011) 5337.
\bibitem{yg1} Y. Grandati, {\it Ann. Phys.} \textbf{326} 2074 (2011); ibid \textbf{327} 185 (2012); \textbf{327} 2411 (2012).
\bibitem{dim} R. K. Yadav et al., {\it arXiv}: 1412.5445 [quant-ph].
\bibitem{bpm5} B Basu-Mallick, B P Mandal and P Roy,{\it arXiv}: 1609.05781 ( To appear in
Annals of Physics)


\bibitem{pdm} B. Midya and B. Roy, {\it Phys. Lett. A} \textbf{373} 4117 (2009).
\bibitem{nfold2} B. Midya, B. Roy, and T. Tanaka, {\it J. Phys. A} \textbf{45} 205303 (2012).
\bibitem{fplank} C.-L. Ho, {\it Ann. Phys.} \textbf{326} 797 (2011).
\bibitem{codext1} D. Dutta and P. Roy, {\it J. Math. Phys.} \textbf{52} 122107 (2011).
\bibitem{codext2} G. Junker and P. Roy, {\it Ann. Phys.} \textbf{270} 155 (1998).
\bibitem{qscat}  R. K. Yadav, A. Khare and B. P. Mandal, {\it Annals of Physics} \textbf {331} (2013) 313; {\it Phys. Lett. B} \textbf{723} (2013) 433; {\it Phys. Lett. A} \textbf{379} (2015) 67.
\bibitem{scatpt}  N. Kumari, R. K. Yadav, A. Khare, B. Bagchi, B. P. Mandal,
{\it Annals of Physics} {\bf 373} (2016) 163.

\bibitem{tdse} A. Schulze-Halberg and B. Roy, {\it J. Math. Phys.} \textbf{55} (2014) 123506.

\bibitem{gtextd}  R. K. Yadav, N. Kumari, A. Khare and B. P. Mandal, {\it Annals of Physics} \textbf {359} (2015) 46.
\bibitem{para_sym} R. K. Yadav, A. Khare, B. Bagchi,  N. Kumari, B. P. Mandal,
{\it J. Math. Phys.} {\bf 57} (2016) 062106-1.

 

\bibitem{cq_12} C. Quesne, {\it SIGMA}  {\bf 8} (2012) 080.



\end{thebibliography}
\end{document}